\documentclass{JHEP3}
\usepackage{graphicx,amsmath,amssymb}
\usepackage{epsfig,multicol}
\usepackage{epsf}
\usepackage{epstopdf}
\input{epsf.sty}

\usepackage{graphicx}
\usepackage{amssymb}
\usepackage{amsmath}

\def\p{\partial}

\def\half{{1\over 2}}
\def\({\left(}
\def\){\right)}
\def\[{\left[}
\def\]{\right]}

\def\e{\begin{equation}}
\def\q{\end{equation}}
\def\m{\begin{eqnarray}}
\def\n{\end{eqnarray}}

\title{Spectral index and running of $g_{NL}$ from an isocurvature scalar field}
\author{Qing-Guo Huang \footnote{huangqg@itp.ac.cn}
\\\small{\em
Kavli Institute for Theoretical Physics China (KITPC), Key Laboratory of Frontiers in Theoretical Physics, Institute of
Theoretical Physics, Chinese Academy of Sciences, Beijing 100190,
China} }

\abstract{
It is possible that the primordial non-Gaussianity is dominated by the higher order terms, such as that set by $g_{NL}$, not $f_{NL}$. In this paper we re-derive the spectral index and work out the running of $g_{NL}$ from a single isocurvature scalar field. The scale dependences of non-Gaussianity parameters are detectable if the mass of isocurvature field is not too small compared to the Hubble parameter during inflation. In addition, we also apply our results to investigate the curvaton model with near quadratic potential in detail. 
}


\begin{document}

\section{Introduction}

Non-Gaussianity \cite{Bartolo:2004if} has become a very important probe into the physics in the early universe. A well-understood non-Gaussianity has a local form which says that the curvature perturbation can be expanded to the  non-linear orders at the same spatial point 
\m
\zeta({\bf x})=\zeta_g({\bf x})+{3\over 5}f_{NL} \zeta_g^2({\bf x})+{9\over 25}g_{NL}\zeta_g^3({\bf x})+...\ ,
\n
where $f_{NL}$ and $g_{NL}$ are the non-Gaussianity parameters which set the sizes of bispectrum and trispectrum respectively. Single-field inflation predicts $f_{NL}\sim {\cal O}(n_s-1)$ which is constrained to be much less than unity. A convincing detection of local form non-Gaussianity will rule out all single-field inflation models.

A large local form non-Gaussianity can be generated by the isocurvature field(s) at the end of multi-field inflation \cite{Lyth:2005qk,Sasaki:2008uc,Huang:2009vk} or deep in the radiation-dominant era, such as curvaton model \cite{Enqvist:2001zp,Lyth:2001nq,Moroi:2001ct,Sasaki:2006kq,Huang:2008rj,Enqvist:2008gk,Huang:2008bg,Huang:2008zj,Nakayama:2009ce,Enqvist:2009ww}. In the literatures, the non-Gaussianity parameters are assumed to be scale-independent. However, recently ones found that a scale-independent $f_{NL}$ is not a generic prediction of inflation \cite{Byrnes:2008zy,Byrnes:2009pe,Byrnes:2010ft,Byrnes:2010xd,Huang:2010cy,Riotto:2010nh,Huang:2010es}. \footnote{The scale dependence of $f_{NL}^{equilateral}$ was discussed in \cite{Chen:2005fe}} For simplicity, the scale-dependent $f_{NL}$ and $g_{NL}$ are parameterized as follows
\m
f_{NL}(k)&=&f_{NL}(k_p)\({k\over k_p}\)^{n_{f_{NL}}+\half \alpha_{f_{NL}}\ln {k\over k_p}},\\
g_{NL}(k)&=&g_{NL}(k_p)\({k\over k_p}\)^{n_{g_{NL}}+\half \alpha_{g_{NL}}\ln {k\over k_p}},
\n
where $k_p$ is a pivot scale. If such a scale dependence is not too small, it can be possibly detected by the forthcoming experiments. For example, in \cite{Sefusatti:2009xu}, the authors showed that 
Planck \cite{:2006uk} and CMBPol \cite{Baumann:2008aq} are able to provide a 1-$\sigma$ uncertainty on the spectral index of $f_{NL}$ for local form bispectrum: 
\m
\Delta n_{f_{NL}}\simeq 0.1 {50\over f_{NL}}{1\over \sqrt{f_{sky}}}\quad \hbox{for Planck},
\n
and
\m
\Delta n_{f_{NL}}\simeq 0.05 {50\over f_{NL}}{1\over \sqrt{f_{sky}}}\quad \hbox{for CMBPol},
\n
where $f_{sky}$ is the sky fraction. The effects on the large-scale structure from scale-dependent $f_{NL}$ were discussed in \cite{Becker:2010hx,Shandera:2010ei}. From the theoretical point of view, the leading order of non-Gaussianity can be higher terms, such as that set by $g_{NL}$, not $f_{NL}$. Assuming $f_{NL}=0$, the current observational limit for $g_{NL}$ are $-7.4\times 10^5<g_{NL}<8.2\times 10^5$ (95 $\%$ C.L.) from cosmic microwave background observations \cite{Smidt:2010sv} and  $-3.5\times 10^5<g_{NL}<8.2\times 10^5$ (95 $\%$ C.L.) from large scale structure observations \cite{Desjacques:2009jb}. Planck will reduce the uncertainty of $g_{NL}$ to $\Delta g_{NL}=1.3\times 10^5$ \cite{Desjacques:2009jb}. More studies on fingerprints of the scale-dependent $g_{NL}$ in CMB and large-scale structure are called for in the future. 

A large non-Gaussianity implies a strong interaction between different cuvature perturbation modes. However, it can be generated by the isocurvature field without self-interaction at all. In order to dig out more physics about the isocurvature field, we need some new observables. In \cite{Byrnes:2009pe,Byrnes:2010ft,Byrnes:2010xd,Huang:2010cy}, we found that $f_{NL}$ generated by a free isocurvature field is scale independent and the spectral index and running of $f_{NL}$ are good discriminators to the self-interaction of isocurvature field. We extend our discussions on the scale dependence of $f_{NL}$ in \cite{Huang:2010cy} to $g_{NL}$ in this paper.

Our paper is organized as follows. In Sec.~2, we use the method in \cite{Huang:2010cy} to derive the spectral and running of $g_{NL}$ from a single isocurvature field. In Sec.~3, for an example, we apply our formula to investigate the curvaton model with near quadratic potential.  More discussions are contained in Sec.~4.

\section{The spectral index and running of $g_{NL}$ from an isocurvature scalar field}

In this paper we consider that the curvature perturbation is generated by the quantum fluctuation of an isocurvature field $\sigma$ which slowly rolls down its potential during inflation. Its dynamics is governed by
\m
\dot \sigma\simeq -{V'(\sigma)\over 3H},
\n
where $V'(\sigma)=dV(\sigma)/d\sigma$ and $H$ is the Hubble parameter.

The gravitational dynamics itself introduces important non-linearities, which will contribute to the final non-Gaussianity in the large-scale CMB anisotropies. 
Based on the so-called $\delta N$ formalism \cite{Starobinsky:1986fxa}, the curvature perturbation produced by the isocurvature field can be expanded to the non-linear orders as follows
\e
\zeta(t_f,{\bf x})=N_{,\sigma}(t_f,t_i)\delta\sigma(t_i,{\bf x})+\half N_{,\sigma\sigma}(t_f,t_i){\delta\sigma}^2(t_i,{\bf x})+{1\over 6}N_{,\sigma\sigma\sigma}(t_f,t_i)\delta\sigma^3(t_i,{\bf x})+... \ ,
\label{zeta}
\q
where $N_{,\sigma}$, $N_{,\sigma\sigma}$ and $N_{,\sigma\sigma\sigma}$ are the first, second and third order derivatives of the number of e-folds with respect to $\sigma$ respectively. Here $t_f$ denotes a  final uniform energy density hypersurface and $t_i$ labels any spatially flat hypersurface after the horizon exit of a given mode. Similar to \cite{Huang:2010cy}, $t_i$ is set to be $t_*(k)$ which is determined by $k=a(t_*)H_*$ for a given mode with comoving wavenumber $k$. Therefore the amplitude of the curvature perturbation is given by
\e
\Delta_{\cal R}^2=N_{,\sigma}^2(t_*)\(H_*\over 2\pi\)^2.
\label{pnh}
\q
Gravitational waves were also generated during inflation and the amplitude of its power spectrum takes the form
\m
\Delta_T^2={H_*^2\over \pi^2/2}.
\n
Here we work on the unit of $M_p=1$. The scale dependence of gravitational wave perturbation is measured by $n_T$ which is defined by
\m
n_T\equiv {d\Delta_T^2\over d\ln k}=-2\epsilon_H,
\n
where
\m
\epsilon_H&\equiv&-{\dot H_*\over H_*^2}.
\n
Usually we introduce a new quantity, the tensor-scalar ratio $r_T$, to measure the amplitude of gravitational waves:
\e
r_T\equiv \Delta_T^2/\Delta_{\cal R}^2={8\over N_{,\sigma}^2(t_*)}.
\q
From Eq.~\eqref{zeta}, the non-Gaussianity parameters are given by 
\e
f_{NL}={5\over 6}{N_{,\sigma\sigma}(t_*)\over N_{,\sigma}^2(t_*)}.
\q
and
\m
g_{NL}={25\over 54}{N_{,\sigma\sigma\sigma}(t_*)\over N_{,\sigma}^3(t_*)}.
\n

For working out the scale dependence of $g_{NL}$, we follow the method in \cite{Huang:2010cy} and introduce a new time $t_r (>t_*)$ which is chosen as a time soon after all the modes of interest exit the horizon during inflation. The value of $\sigma$ at $t_r$ is related to that at time $t_*$ by
\e
\int_{\sigma_*}^{\sigma_r}{d\sigma\over V'(\sigma)}=-\int_{t_*}^{t_r} {dt\over 3H(t)}.
\q
Therefore we have
\m
\left. {\p \sigma_r\over \p \sigma_*} \right|_{t_*}&=&{V'(\sigma_r)\over V'(\sigma_*)}, \\
\left. {\p \sigma_r\over \p t_*} \right|_{\sigma_*}&=&{V'(\sigma_r)\over 3H(t_*)}.
\n
Considering
\m
{d\over dt_*}F(\sigma_r)={\p F(\sigma_r)\over \p\sigma_r}(\dot \sigma_* {\p \sigma_r\over \sigma_*}+{\p \sigma_r\over \p t_*})
\n
and $3H\dot \sigma_*=-V'(\sigma_*)$, one finds 
\e
{d\over d\ln k}F(\sigma_r)={d\over H_*dt_*}F(\sigma_r)=0,
\q
which implies that $F(\sigma_r)$ is scale independent.
Taking into account that $\sigma_r$ is a function of $\sigma_*$, we have
\m
N_{,\sigma}(t_*)&=&{\p\sigma_r\over \p\sigma_*} {\p N(\sigma_r)\over \p\sigma_r},\\
N_{,\sigma\sigma}(t_*)&=&{\p^2\sigma_r\over \p\sigma_*^2} {\p N(\sigma_r)\over \p\sigma_r}+\({\p\sigma_r\over \p\sigma_*}\)^2  {\p^2 N(\sigma_r)\over \p\sigma_r^2},\\
N_{,\sigma\sigma\sigma}(t_*)&=&{\p^3\sigma_r\over \p\sigma_*^3} {\p N(\sigma_r)\over \p\sigma_r}+3{\p\sigma_r\over \p\sigma_*}{\p^2\sigma_r\over \p\sigma_*^2} {\p^2 N(\sigma_r)\over \p\sigma_r^2}+\({\p\sigma_r\over \p\sigma_*}\)^3  {\p^3 N(\sigma_r)\over \p\sigma_r^3}.
\n
Since ${\p N(\sigma_r)\over \p\sigma_r}$, ${\p^2 N(\sigma_r)\over \p\sigma_r^2}$ and ${\p^3 N(\sigma_r)\over \p\sigma_r^3}$ are scale independent, one obtains
\m
{d\ln N_{,\sigma}(t_*)\over d\ln k}&=&\eta_{\sigma\sigma},\\
{d\ln N_{,\sigma\sigma}(t_*)\over d\ln k}&=&2\eta_{\sigma\sigma}+\eta_3 {N_{,\sigma}(t_*)\over N_{,\sigma\sigma}(t_*)},\\
{d\ln N_{,\sigma\sigma\sigma}(t_*)\over d\ln k}&=&3\eta_{\sigma\sigma}+3\eta_3 {N_{,\sigma\sigma}(t_*) \over N_{,\sigma\sigma\sigma}(t_*)}+\xi_4 {N_{,\sigma}(t_*) \over N_{,\sigma\sigma\sigma}(t_*)},
\n
where the slow-roll equation for $\sigma$ is adopted and
\m
\eta_{\sigma\sigma}\equiv {V''(\sigma_*)\over 3H_*^2},\quad
\eta_3\equiv {V'''(\sigma_*)\over 3H_*^2},\quad
\xi_4={V^{(4)}(\sigma_*)\over 3H_*^2}.
\n
From the above results, the spectral index of $\Delta_{\cal R}^2$, $f_{NL}$ and $g_{NL}$ are respectively given by
\m
n_s\equiv 1+{d\ln \Delta_{\cal R}^2\over d\ln k}=1+2\eta_{\sigma\sigma}-2\epsilon_H,
\label{ns}
\n
\m
n_{f_{NL}}\equiv {d\ln |f_{NL}|\over d\ln k}=\eta_3 {N_{,\sigma}(t_*)\over N_{,\sigma\sigma}(t_*)},
\n
and
\m
n_{g_{NL}}&\equiv& {d\ln |g_{NL}|\over d\ln k}=3\eta_3 {N_{,\sigma\sigma}(t_*)\over N_{,\sigma\sigma\sigma}(t_*)}+\xi_4 {N_{,\sigma}(t_*)\over N_{,\sigma\sigma\sigma}(t_*)}, \\
&=&2 {f_{NL}^2\over g_{NL}} n_{f_{NL}}+{25\over 432}\xi_4{r_T\over g_{NL}}.
\n
Our results are the same as those in \cite{Byrnes:2010ft}. The scale dependence of the non-Gaussianity parameters from an isocurvature field is proportional to third and fourth order derivatives of its potential. Therefore the spectral indices of $f_{NL}$ and $g_{NL}$ are really the good discriminators to measure the self-interaction of such an isocurvature scalar field.

In fact the indicies $n_s$ and $n_T$ may be scale dependent as well. Their scale dependences are measured by the so-called runnings which are defined by
\m
\alpha_s\equiv {dn_s\over d\ln k},\quad \hbox{and}\quad \alpha_T\equiv {dn_T\over d\ln k}. 
\n
The dynamics of slow-roll inflation is governed by the inflaton field $\phi$ whose potential is denoted as $V_I(\phi)$ and then we have 
\m
{d\epsilon_H \over d \ln k}=-2\epsilon_H\eta_{\phi\phi}+4\epsilon_H^2,
\n
where
\m
\eta_{\phi\phi}\equiv {V_I''(\phi)\over V_I(\phi)}. 
\n
Considering 
\m
{d\eta_{\sigma\sigma} \over d\ln k}=2\epsilon_H\eta_{\sigma\sigma}-\xi_3,
\n
we obtain
\m
\alpha_T&=&4\epsilon_H\eta_{\phi\phi}-8\epsilon_H^2,\\
\alpha_s&=&\alpha_T+4\epsilon_H \eta_{\sigma\sigma}-2\xi_3. 
\n
where 
\m
\xi_3={V'V'''\over 9H_*^4}.
\n
In the future, the spectral index and running of curvature perturbation might be measured precisely. There might be an opportunity to measure $n_T$ by CMBPol \cite{Baumann:2008aq} as long as $r_T$ is not too small \cite{Zhao:2011zb}.  One can imagine that it is quite hard to measure $\alpha_T$. From the above formula, $\eta_{\phi\phi}$ makes contribution to $\alpha_s$ as well. So it is difficult to re-construct the potential of isocurvature field from $n_s$ and $\alpha_s$.

In \cite{Huang:2010cy} the running of spectral index $n_{f_{NL}}$ from an isocurvature field is derived as follows
\m
\alpha_{f_{NL}}\equiv {dn_{f_{NL}}\over d\ln k}=(2\epsilon_H-\eta_{\sigma\sigma}-\eta_4)n_{f_{NL}}-n_{f_{NL}}^2,
\label{alphafnl}
\n
where
\m
\eta_4\equiv {V'V^{(4)}\over 3H^2V'''}.
\n
Similarly, we define the running of $n_{g_{NL}}$, namely
\e
\alpha_{g_{NL}}\equiv {dn_{g_{NL}}\over d\ln k}.
\q
Taking into account
\m
{d\xi_4\over d\ln k}=(2\epsilon_H-\xi_5)\xi_4
\n
and
\m
{dr_T\over d\ln k}=-2r_T\eta_{\sigma\sigma},
\n
the running of $g_{NL}$ becomes
\m
\alpha_{g_{NL}}=&-&n_{g_{NL}}(2\eta_{\sigma\sigma}-2\epsilon_H)-n_{g_{NL}}^2\nonumber \\
&+&{1\over g_{NL}} \[2f_{NL}^2n_{f_{NL}}(\eta_{\sigma\sigma}-\eta_4+n_{f_{NL}})
-{25\over 432} \xi_4 r_T \xi_5\],
\label{alphagnl}
\n
where
\m
\xi_5={V'V^{(5)}\over 3H_*^2V^{(4)}}.
\n
If $n_{g_{NL}}\sim {\cal O}(1)$, $\alpha_{g_{NL}}$ is roughly the same order of $n_{g_{NL}}$, and hence $n_{g_{NL}}$ is not a reliable quantity to characterize the scale dependence of $g_{NL}$ any more. Actually there is one another non-Gaussianity parameter $\tau_{NL}$ which measures a special local form  trispectrum. In our setup, $\tau_{NL}$ is not an independent parameter and it is related to $f_{NL}$ by $\tau_{NL}=({6\over 5}f_{NL})^2$. Therefore $n_{\tau_{NL}}=2n_{f_{NL}}$ and $\alpha_{\tau_{NL}}=2\alpha_{f_{NL}}$.

Combining with Eq.~(\ref{ns}), the running of spectral index of $g_{NL}$ becomes
\m
\alpha_{g_{NL}}=&-&n_{g_{NL}}(n_s-1+n_{g_{NL}})\nonumber \\
&+&{1\over g_{NL}} \[2f_{NL}^2n_{f_{NL}}(\eta_{\sigma\sigma}-\eta_4+n_{f_{NL}})
-{25\over 432} \xi_4 r_T \xi_5\]
\n
For the special case with $f_{NL}=0$, the spectral index and running of $g_{NL}$ are simplified to be
\m
n_{g_{NL}}&=&{25\over 432} \xi_4{r_T\over g_{NL}},\\
\alpha_{g_{NL}}&=&n_{g_{NL}}(1-n_s-\xi_5-n_{g_{NL}}).
\n
Now for an isocurvature field with renormalizable potential $V(\sigma)\sim \sigma^n$, where $n$ is an integer not larger than 4, $\xi_5=0$ and we obtain a consistency relation
\m
\alpha_{g_{NL}}=n_{g_{NL}}(1-n_s-n_{g_{NL}}).
\n
We hope that the running of $g_{NL}$ can be detected as well even though it is expected to be very difficult in the near future.


For an instance, we consider an isocurvature field $\sigma$ who has a polynomial potential,
\m
V(\sigma)=\half m^2\sigma^2+\lambda m^4 \({\sigma\over m}\)^n.
\label{pt}
\n
Similar to \cite{Enqvist:2008gk,Huang:2008bg,Huang:2010cy}, we introduce a new parameter to measure its self-interaction term compared to its mass term as follows
\m
s\equiv 2\lambda\({\sigma_*\over m}\)^{n-2}.
\n
Here $s>-2/n$. Otherwise the isocurvature field will run away to the infinity. 
Accordingly, $\eta_{\sigma\sigma}$, $\eta_4$ and $\xi_4$ are given by 
\m
\eta_{\sigma\sigma}=\eta_{mm}\(1+{n(n-1)\over 2}s\)
\label{etasm}
\n
and
\m
\eta_3={n(n-1)(n-2)\over 2}\eta_{mm}{s\over \sigma_*},\quad \xi_4={n(n-1)(n-2)(n-3)\over 2}\eta_{mm}{s\over \sigma_*^2},
\n
where $\eta_{mm}={m^2/ 3H_*^2}$. Here $m^2$ is assumed to be positive. The spectral indices of $f_{NL}$ and $g_{NL}$ are 
\m
n_{f_{NL}}&=& \hbox{sign}(N_{,\sigma})  {5n(n-1)(n-2)\over 24\pi \Delta_{\cal R}} \eta_{mm} {H_*\over \sigma_*}{s\over f_{NL}},\\
n_{g_{NL}}&=&\hbox{sign}(N_{,\sigma}) {5n(n-1)(n-2)\over 12\pi \Delta_{\cal R}}\eta_{mm}{H_*\over \sigma_*}{f_{NL}\over g_{NL}} s \nonumber \\
&+& {25n(n-1)(n-2)(n-3)\over 432\pi^2\Delta_{\cal R}^2} \eta_{mm}{H_*^2\over \sigma_*^2}{1\over g_{NL}}s. 
\label{dd}
\n
If $|f_{NL}|<{5(n-3)\over 18 \Delta_{\cal R}}{H_*/2\pi\over \sigma_*}$, the second term on the right hand side of Eq.~\eqref{dd} becomes dominant. Since both $n_{f_{NL}}$ and $n_{g_{NL}}$ are proportional to $\eta_{mm}$, the scale dependences of $f_{NL}$ and $g_{NL}$ are detectable only when the mass of isocurvature field is not too small compared to the Hubble parameter during inflation.

\begin{figure}[h]
\begin{center}
\includegraphics[width=14cm]{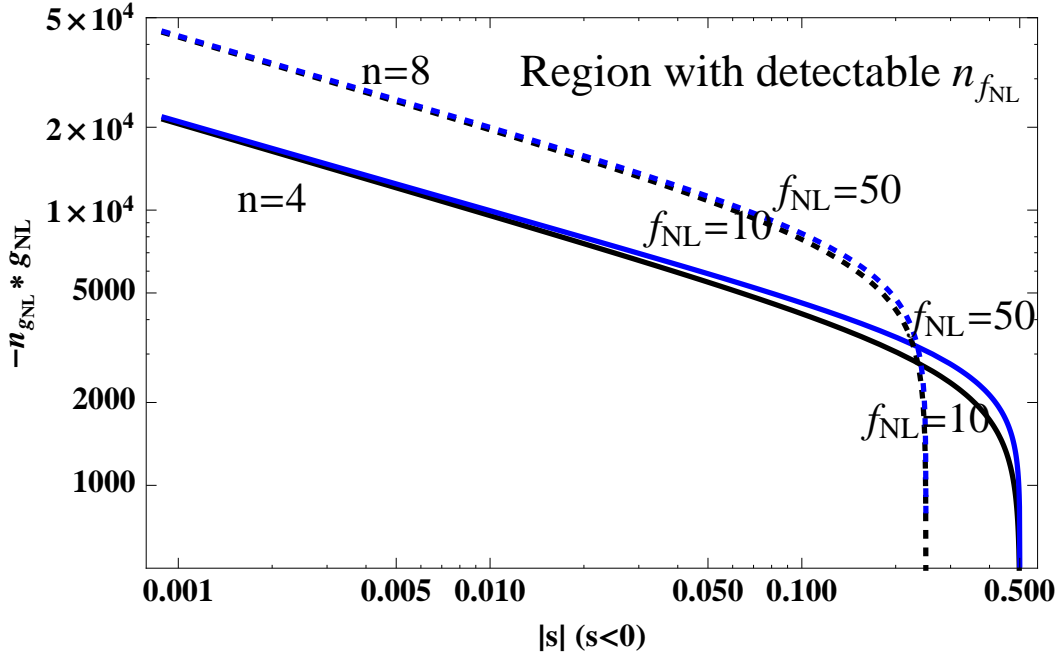}\\
\vspace{5mm}
\includegraphics[width=14cm]{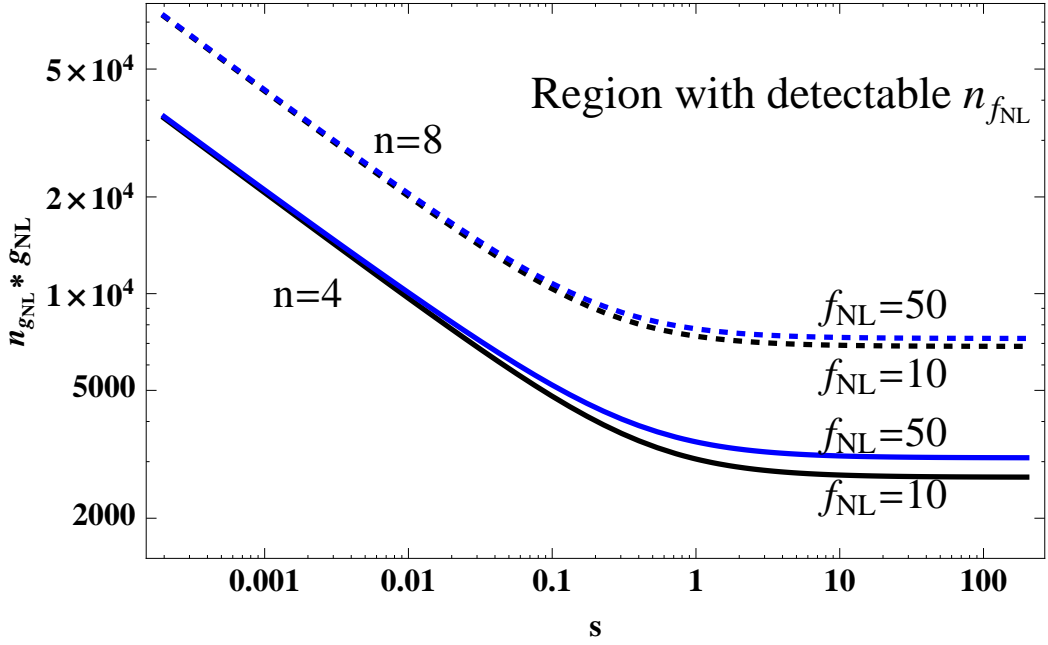}
\end{center}
\caption{The regions of $n_{g_{NL}} \cdot g_{NL}$ above the curves correspond to the cases with a detectable $n_{f_{NL}}$.  }
\label{fig:ngnlgnl}
\end{figure}

Because the Compton wavelength of isocurvature field is large compared to the Hubble size during inflation, its quantum fluctuations leads to a typical vacuum expectation value of $\sigma$ \cite{Huang:2010cy}, i.e. 
\e
\sigma_*=\sqrt{3\over 8\pi^2(1+{n\over 2}s)}{H_*^2\over m}. 
\q
Taking the above result into account, $n_{f_{NL}}$ and $n_{g_{NL}}$ are simplified to be 
\m
n_{f_{NL}}&=&\hbox{sign}(N_{,\sigma})\cdot 2.3\times 10^3 n(n-1)(n-2) {1\over f_{NL}} \({m \over H_*}\)^3 s \sqrt{1+{n\over 2}s},\\
n_{g_{NL}}&=&\hbox{sign}(N_{,\sigma})\cdot 4.6\times 10^3 n(n-1)(n-2){f_{NL}\over g_{NL}} \({m \over H_*}\)^3 s \sqrt{1+{n\over 2}s} \nonumber \\
&+& 2.1\times 10^7 n(n-1)(n-2)(n-3) {1\over g_{NL}} \({m\over H_*}\)^4 s (1+{n\over 2}s). 
\n
Here we adpot WMAP normalization: $\Delta_{\cal R}=4.96\times 10^{-5}$ \cite{Komatsu:2010fb}. For a model with detectable $n_{f_{NL}}$ by PLANCK, $|n_{f_{NL}}\cdot f_{NL}|\gtrsim 5$ \cite{Sefusatti:2009xu} which implies 
\m
{m\over H_*}\gtrsim 0.13 \[n(n-1)(n-2)|s| \sqrt{1+{n\over 2}s}\]^{-1/3}.
\label{cons}
\n
For the model with $\hbox{sign}(N_{,\sigma})>0$, the region of $n_{g_{NL}}\cdot g_{NL}$ with detectable $n_{f_{NL}}$ is illustrated in Fig.~\ref{fig:ngnlgnl}.
Because $\sigma$ is considered to be a general isocurvature field, one cannot figure out the value of $f_{NL}$ only from $m/H_*$ and $s$. The current constraints on $f_{NL}$ from WMAP \cite{Komatsu:2010fb} is $f_{NL}=32\pm 21$. So, for example, we consider $f_{NL}=10$ and $50$ which correspond to the black and blue curves in Fig.~\ref{fig:ngnlgnl}, respectively. We find that the lower bound on $|n_{g_{NL}}\cdot g_{NL}|$ is not sensitive to the value of $f_{NL}$. On the other hand, the mass of isocurvature field is assumed to be smaller than the Hubble parameter during inflation and then Eq.~\eqref{cons} implies that $|s|$ cannot be too small.


\section{$n_{g_{NL}}$ in the curvaton model with near quadratic potential}

We start with the curvaton whose potential takes the form in Eq.~\eqref{pt}. In this section, we focus on the case in which the self-interaction term is much smaller than the mass term. For $|s|\ll 2/n$ and $m^2>0$, the amplitude of scalar power spectrum, $f_{NL}$ and $g_{NL}$ are respectively given by
\m
\Delta_{\cal R}&=& {2\over 3}f_D q {H_*/2\pi\over \sigma_*},\label{deltas} \\
f_{NL}&=&{5\over 4f_D}(1+h_2)-{5\over 3}-{5f_D\over 6},\label{fnl}\\
g_{NL}&=&{25\over 54}\[{9\over 4f_D^2}(h_3+3h_2)-{9\over f_D}(1+h_2)+\half (1-9h_2)+10f_D+3f_D^2\],\label{gnl}
\n
where $f_D={3\Omega_{\sigma,D}\over 4-\Omega_{\sigma,D}}$, and
\m
q&=&{w(x_0)+n(n-1)g(n,x_0)s/2\over w(x_0)+n g(n,x_0)s/2},\\
h2&=&{w(x_0)+n g(n,x_0)s/2\over (w(x_0)+n(n-1)g(n,x_0)s/2)^2}n(n-1)(n-2)g(n,x_0)s/2,\\
h3&=&{(w(x_0)+n g(n,x_0)s/2)^2\over (w(x_0)+n(n-1)g(n,x_0)s/2)^3}n(n-1)(n-2)(n-3)g(n,x_0)s/2.
\n
See, for example, \cite{Huang:2008bg} in detail.
Here $\Omega_{\sigma,D}$ is the fraction of curvaton energy density in the total energy density budget at the time of its decay,
\m
w(x_0)=2^{1/4} \Gamma(5/4)x_0^{-1/4}J_{1/4}(x_0),
\n
and
\m
g(n,x_0)=&&\pi 2^{(n-5)/4}\Gamma(5/4)^{n-1} x_0^{-1/4}\nonumber \\
&&\times \[J_{1/4}(x_0)\int_0^{x_0} J_{1/4}^{n-1}(x) Y_{1/4}(x)x^{(6-n)/4}dx \right. \nonumber \\
&&\left. -Y_{1/4}(x_0)\int_0^{x_0}J_{1/4}^n(x)x^{(6-n)/4}dx\],
\n
and $x_0=mt_0=1$ denotes the time when curvaton starts to oscillate. Taking Eq.~\eqref{deltas} into account, the tensor-scalar ratio becomes
\m
r_T={18\over f_D^2 q^2}\sigma_*^2.
\n
For a sub-Planckian value of $\sigma_*$, the tensor-scalar ratio is much smaller than one if $f_D$ is not too small.

\begin{figure}[h]
\begin{center}
\includegraphics[width=7.5cm]{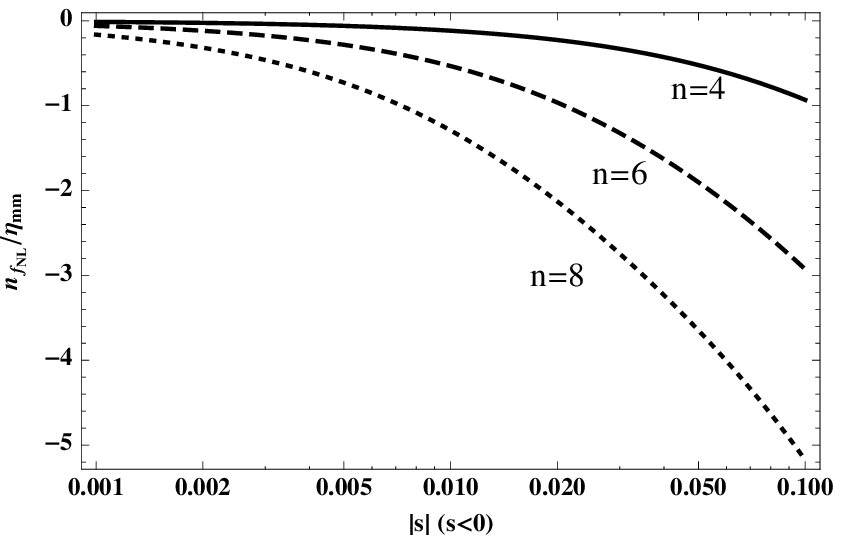}\ \includegraphics[width=7.5cm]{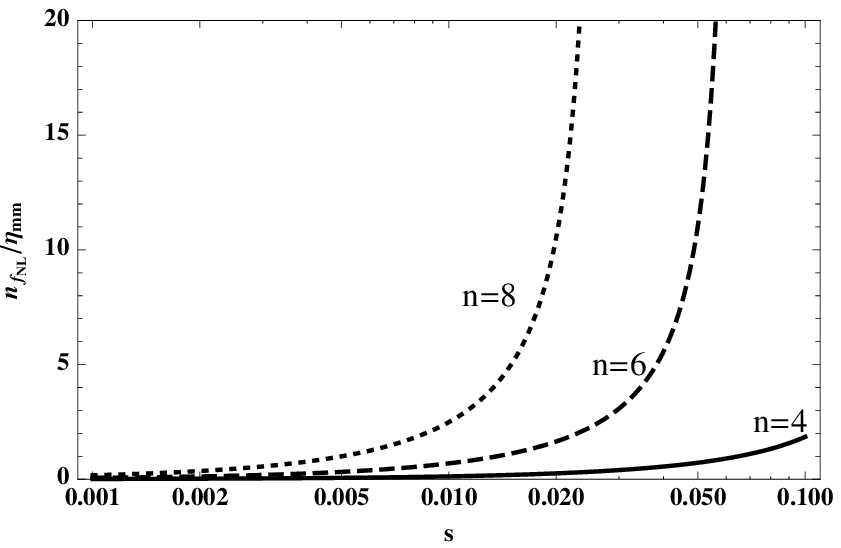}\\
\includegraphics[width=7.5cm]{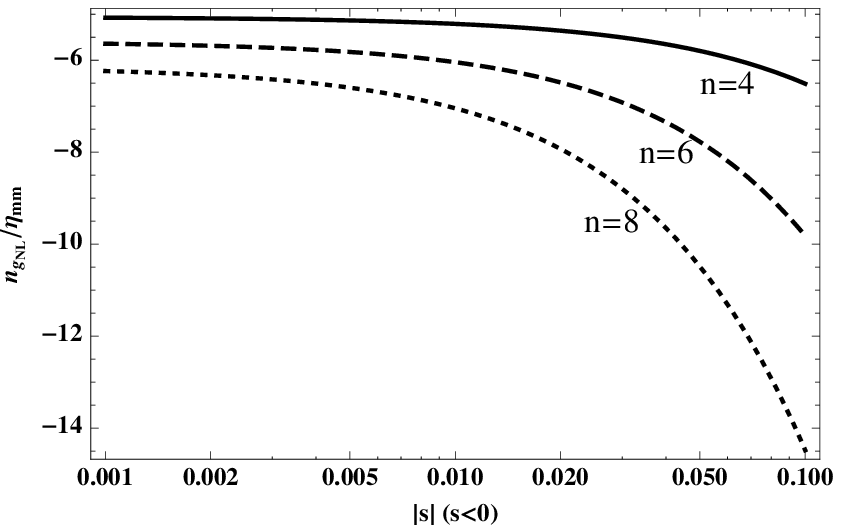}\ \includegraphics[width=7.5cm]{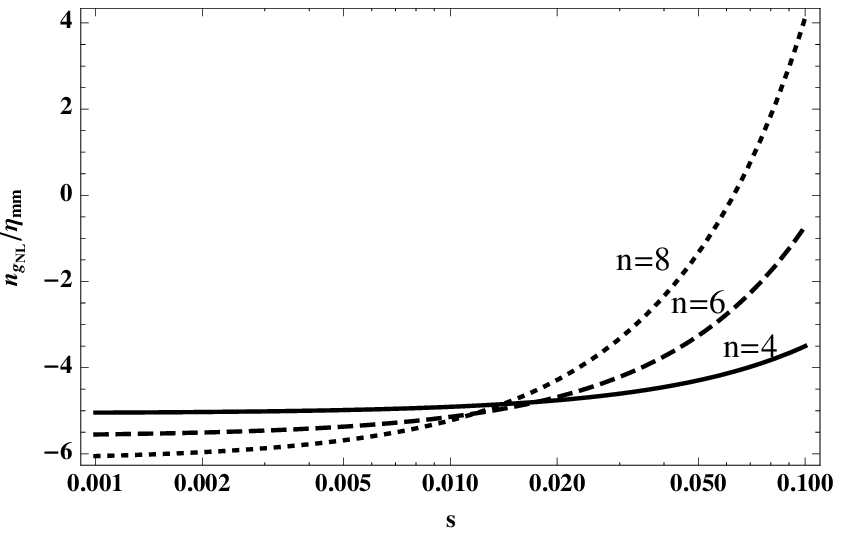}
\end{center}
\caption{The values of $n_{f_{NL}}/\eta_{mm}$ and $n_{g_{NL}}/\eta_{mm}$ in the curvaton model with a polynomial potential. The solid, dashed and dotted lines correspond to $n=4,\ 6,\ 8$ respectively.  }
\label{fig:gngnl}
\end{figure}

In order to obtain a large non-Gaussianity, $f_D$ should be smaller than one. For the curvaton model with near quadratic potential, the non-Gaussianity parameters are approximately given by
\m
f_{NL}&\simeq&{5\over 4f_D}(1+h_2),\\
g_{NL}&\simeq&{25\over 24 f_D^2}(h_3+3h_2).
\n
One can easily calculate the spectral indices of $f_{NL}$ and $g_{NL}$:
\m
n_{f_{NL}}&\simeq& {n(n-1)(n-2)\over 2}\eta_{mm}{s\over q(1+h_2)},\\
n_{g_{NL}}&\simeq& {n(n-1)(n-2)\over 2}\eta_{mm}{3 s\over q^2(h_3+3h_2)} \[q(1+h_2)+{n\over 3}-1\].
\n
The spectral indices of both $f_{NL}$ and $g_{NL}$ are independent on $f_D$. Here $n_{f_{NL}}$ and $n_{g_{NL}}$ are illustrated in Fig.~\ref{fig:gngnl} for $s<0$ and $s>0$ respectively.  
In this setup there are four model-dependent parameters: $n$, $s$, $\eta_{mm}$ and $f_D$. They can be fixed by four observables: $f_{NL}$, $n_{f_{NL}}$, $g_{NL}$ and $n_{g_{NL}}$. The spectral indices of $f_{NL}$ and $g_{NL}$ are very useful for us to re-construct curvaton potential.

\begin{figure}[h]
\begin{center}
\includegraphics[width=7.5cm]{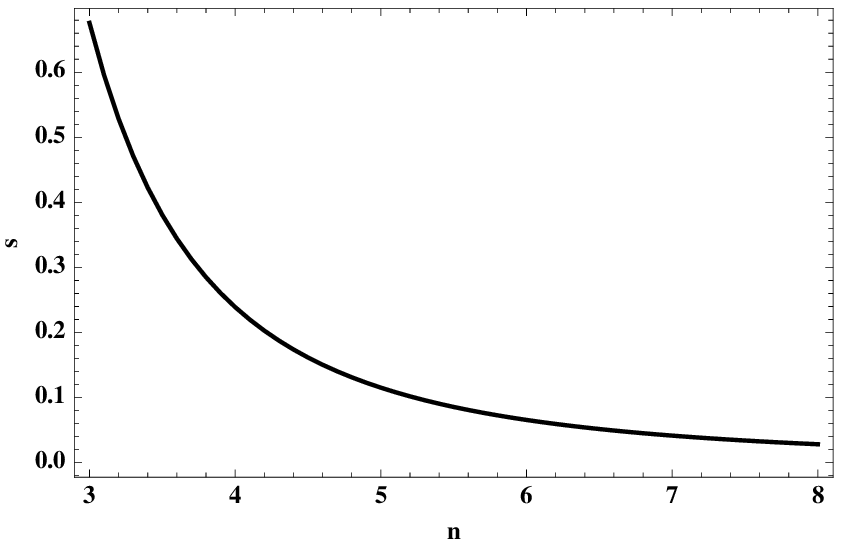}\ \includegraphics[width=7.5cm]{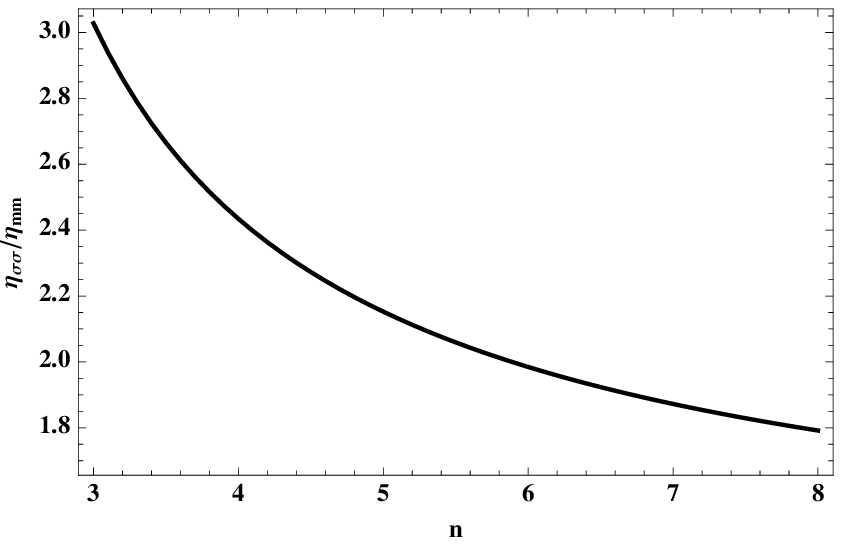}\\
\includegraphics[width=7.5cm]{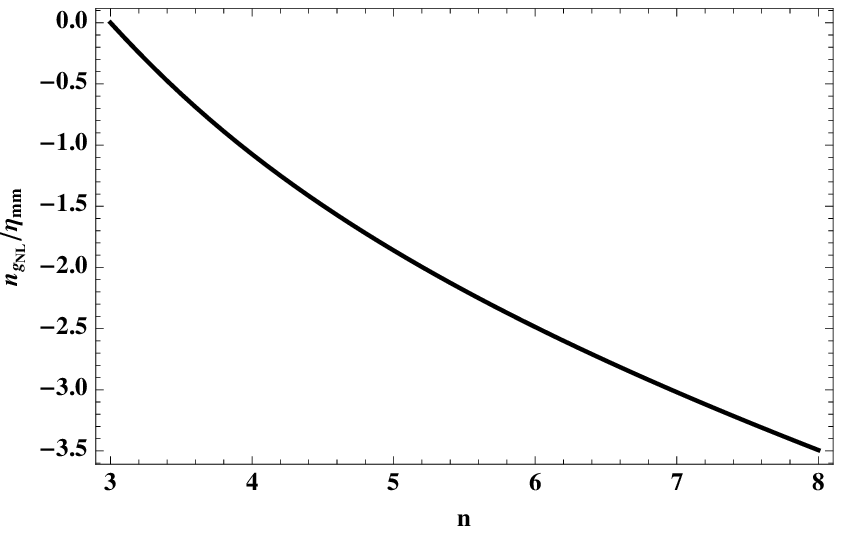}\ \includegraphics[width=7.5cm]{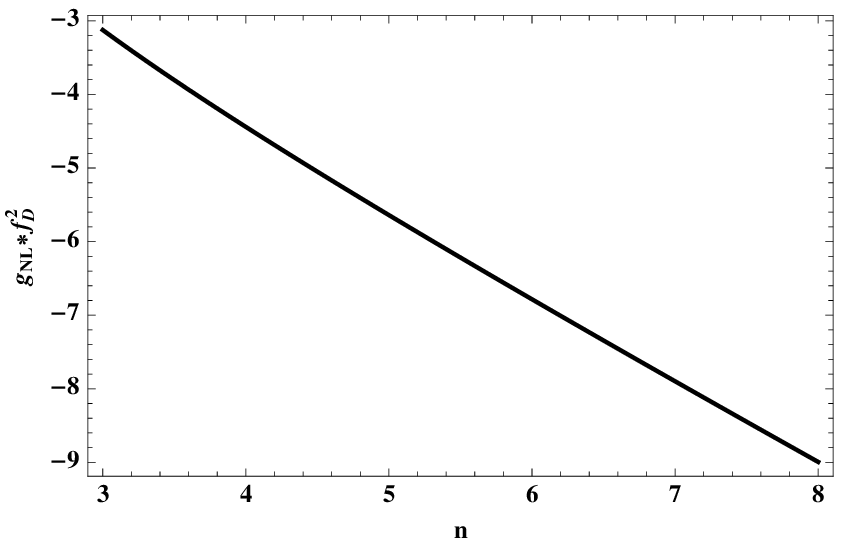}
\end{center}
\caption{The values of $s$, $\eta_{\sigma\sigma}/\eta_{mm}$, $n_{g_{NL}}/\eta_{mm}$ and $g_{NL}*f_D^2$ in the curvaton model with $f_{NL}=0$.  }
\label{fig:ngnl}
\end{figure}

We also notice that $f_{NL}$ can be tuned to be zero even when $f_D\ll 1$ for the curvaton model with near quadratic potential. In this case the higher order terms, such as $g_{NL}$, still leads to a large non-Gaussianity which can be potentially detected by the forthcoming observations as well. Now $h_2=-1$, but $g_{NL}$ can still be quite large 
\m
g_{NL}\simeq {25\over 24}{h_3-3\over f_D^2}.
\n
The spectral index of $g_{NL}$ is simplified to be 
\m
n_{g_{NL}}=-\half n(n-1)(n-2)(n-3)\eta_{mm}{s\over q^2(3-h_3)}.
\label{cngnl}
\n
Here, for a given $n$, $s$ can be fixed by the condition $h_2=-1$. Our numerical results are summarized in Fig.~\ref{fig:ngnl}.
From Eq.~\eqref{cngnl}, $n_{g_{NL}}$ is proportional to $\eta_{mm}$. Larger $\eta_{mm}$, larger $n_{g_{NL}}$. On the other hand, if $r_T$ and $n_T$ are not too small, they might be measured in the near future \cite{Zhao:2011zb}, and then $\eta_{\sigma\sigma}$ and $\eta_{mm}$ can be fixed by $\eta_{\sigma\sigma}=(n_s-1-n_T)/2$ and Eq.~\eqref{etasm}. The order of magnitude of both $\eta_{\sigma\sigma}$ and $\eta_{mm}$ is expected to be ${\cal O}(10^{-2})$. Therefore the spectral index of $g_{NL}$ in the curvaton model with $f_{NL}=0$ is negative and its order of magnitude is roughly $-{\cal O}(10^{-2})$.

\section{Discussions}

Inflation is driven by the vacuum energy of inflaton field. The vacuum energy density is almost a constant with the expansion of universe and then the Hubble parameter almost did not decrease during inflation. However it is not an exact constant, but slowly decreased. WMAP data implies that the power spectrum of curvature perturbation is just near scale invariant $(n_s=0.968\pm0.012)$, not exactly scale invariant. The tilt of power spectrum comes from the fact that the inflaton potential is not exactly flat.

A large local form non-Gaussianity can be naturally generated by an isocurvature field on the super-horizon scales. If the isocurvature field is a free field (without self-interaction), the non-Gaussianity parameters, such as $f_{NL}$ and $g_{NL}$, should be scale independent. More generally, one may expect that the isocurvature field self-interacts with itself and then the non-Gaussianity parameters also depend on the wavelengths of perturbation modes. On the other hand, one can learn how the isocurvature field interacts with itself from the scale dependences of the non-Gaussianity parameters.

In this paper we derive the spectral index and running of $g_{NL}$ from a general isocurvature field. The typical region of $n_{g_{NL}}$ for the model with detectable $n_{f_{NL}}$ is illustrated in Fig.~\ref{fig:ngnlgnl}. Applying our results to the curvaton model with near quadratic potential, we find that both $n_{f_{NL}}$ and $n_{g_{NL}}$ are independent on $f_D$. Therefore one can usually tune the free parameter $f_D$ to achieve a large non-Gaussianity. However, in curvaton model, one can tune $f_{NL}$ to be zero even when $f_D\ll 1$, but $g_{NL}\sim 1/f_D^2$ is still large. In this special case, $n_{g_{NL}}\sim -{\cal O}(10^{-2})$.

In the literatures, the near scale-invariant variables are usually re-parametrized by 
\m
O(k)=O(k_p)\({k\over k_p}\)^{n+\half \alpha\ln {k\over k_p}}. 
\n
This expression is reliable around $k=k_p$. Actually one can re-write the above formula as follows
\m
O(k)=O(k_p)\[1+n\ln{k\over k_p}+\half {\tilde \alpha} \(\ln {k\over k_p}\)^2+...\],
\n
where $\tilde \alpha=n^2+\alpha$ is the new ``running" of spectral index $n$. If we adopt this new definition, the terms $-n_{f_{NL}}^2$ and $-n_{g_{NL}}^2$ on the right hand sides of $\alpha_{f_{NL}}$ and $\alpha_{g_{NL}}$ in Eqs. \eqref{alphafnl} and \eqref{alphagnl} can be absorbed into ${\tilde \alpha}_{f_{NL}}$ and ${\tilde \alpha}_{g_{NL}}$ respectively. An advantage of this new definition is that ${\tilde \alpha}_{f_{NL}}$ and ${\tilde \alpha}_{g_{NL}}$ are respectively much smaller than $n_{f_{NL}}$ and $n_{g_{NL}}$ even when $n_{f_{NL}},\ n_{g_{NL}}\sim {\cal O}(1)$.

Once the large local form non-Gaussianity is confirmed by the forthcoming observations, the scale dependence of the non-Gaussianity parameters will be the next important issue. How to search for the signal of scale dependence of $g_{NL}$ in the CMB and large-scale structure data is still needed to be done in the near future.

\vspace{1.4cm}

\noindent {\bf Acknowledgments}

\vspace{.5cm}

QGH is supported by the project of Knowledge Innovation
Program of Chinese Academy of Science and a grant from NSFC.




\end{document}